\documentclass[sigconf, nonacm]{acmart}

\DeclareUnicodeCharacter{2212}{\ensuremath{-}}
\usepackage[a-2b]{pdfx}
\usepackage{hyperref} 

\usepackage[utf8]{inputenc}
\hypersetup{pdfencoding=unicode}
\inputencoding{utf8}
\makeatother

\usepackage{natbib}
\usepackage{graphicx}
\usepackage{amsthm}
\usepackage{amsmath} 
\usepackage{amsfonts} 

\newtheorem{problem}{Problem Definition}
\usepackage{comment}
\usepackage{algorithm}
\usepackage[noend]{algpseudocode}
\usepackage{subcaption}
\usepackage{color}
\usepackage[labelfont=bf,textfont=normalfont,singlelinecheck=off,
justification=centering]{subcaption}

\usepackage{siunitx}
\algnewcommand\algorithmicforeach{\textbf{for each}}
\algdef{S}[FOR]{ForEach}[1]{\algorithmicforeach\ #1\ \algorithmicdo}
\usepackage{multirow}

\usepackage[group-separator={,},group-minimum-digits=4]{siunitx}
\usepackage{tikz}
\usetikzlibrary{shapes, arrows.meta, positioning, fit, backgrounds, shadows}

\begin{document}
\title{ALER: An Active Learning Hybrid System for Efficient Entity Resolution}

\author{Dimitrios Karapiperis}
\orcid{0000-0002-3878-5988}
\affiliation{%
  \institution{International Hellenic University}
   \city{Thessaloniki}
  \country{Greece}
}
\email{dkarapiperis@ihu.edu.gr}

\author{Leonidas Akritidis}
\orcid{0000-0001-6602-0723}
\affiliation{%
  \institution{International Hellenic University}
   \city{Thessaloniki}
  \country{Greece}
}
\email{lakritidis@ihu.edu.gr}

\author{Panayiotis Bozanis}
\orcid{0000-0001-9435-1829}
\affiliation{%
  \institution{International Hellenic University}
   \city{Thessaloniki}
  \country{Greece}
}
\email{pbozanis@ihu.edu.gr}

\author{Vassilios S. Verykios}
\orcid{0000-0002-9758-0819}
\affiliation{%
  \institution{Hellenic Open University}
   \city{Patras}
  \country{Greece}
}
\email{verykios@eap.gr}

\newcommand{\dv}[1]{\textcolor{blue}{#1}\PackageWarning{DV:}{#1!}}

\begin{abstract}
Entity Resolution (ER) is a critical task for data integration, yet state-of-the-art supervised deep learning models remain impractical for many real-world applications due to their need for massive, expensive-to-obtain labeled datasets. While Active Learning (AL) offers a potential solution to this "label scarcity" problem, existing  approaches introduce severe scalability bottlenecks. Specifically, they achieve high accuracy but incur prohibitive computational costs by re-training complex models from scratch or solving NP-hard selection problems in every iteration. In this paper, we propose \textbf{ALER}, a novel, semi-supervised pipeline designed to bridge the gap between semantic accuracy and computational scalability. ALER eliminates the training bottleneck by using a frozen bi-encoder architecture to generate static embeddings once and then iteratively training a lightweight classifier on top. To address the memory bottleneck associated with large-scale candidate pools, we first select a representative sample of the data and then use K-Means to partition this sample into semantically coherent chunks, enabling an efficient AL loop. We further propose a hybrid query strategy that combines "confused" and "confident" pairs to efficiently refine the decision boundary while correcting high-confidence errors.Extensive evaluation demonstrates ALER's superior efficiency, particularly on the large-scale DBLP dataset: it accelerates the training loop by $1.3\times$ while drastically reducing resolution latency by a factor of $3.8$ compared to the fastest baseline.
\end{abstract}

\maketitle

\section{Introduction}
Entity Resolution (ER) is a fundamental and long-standing challenge in data management. Its goal is to identify records across one or more datasets that refer to the same real-world entity. As a critical step in data integration, ER enables the creation of unified, high-quality knowledge bases from disparate and often noisy data sources. Practical examples range from matching customer records to social media profiles, linking bibliographic records across academic databases, to de-duplicating product catalogs. The core difficulty lies in accurately matching entities despite textual variations like misspellings, abbreviations, and different data formats. Furthermore, the prohibitive quadratic complexity of comparing all possible record pairs makes a naive approach infeasible. To address this, the ER process is almost universally split into two main phases: \textit{Blocking} and \textit{Matching}. The former's purpose is to prune the massive search space. The latter is a more sophisticated phase that performs the final classification (Match/No-Match) on the candidate pairs formulated during blocking.

For decades, efficient ER has been dominated by foundational lexical techniques. These include methods like token blocking~\cite{christen2012data}, which utilizes shared tokens, suffix arrays~\cite{suffix1} used for prefix-based matching,  TF-IDF weighting~\cite{christen2012data, sparkly}, or MinHash shingling \cite{tkde2024}, which creates compact record signatures. While these lexical techniques are efficient at pruning the search space, they share a fundamental weakness: they are blind to the underlying semantic concept. Their reliance on syntactic token overlap makes them brittle and prone to failure when records are semantically similar but lexically distinct. For example, a lexical method would fail to match a user query \textit{"PVLDB"} with \textit{"Proceedings of the VLDB Endowment"}, as they share no meaningful keywords. This semantic gap also leads to a critical failure mode: an inability to distinguish between high similarity and true identity. A lexical approach can easily be confused by highly similar but distinct records—such as an "Apple iPhone 13" versus an "Apple iPhone 14"—and may incorrectly group them. This inability to capture nuanced, real-world meaning necessitates the use of more sophisticated, semantically-aware models.

The development of deep learning, particularly Transformer-based pre-trained language models like BERT \cite{bert} and its sentence-level variant Sentence-BERT (SBERT) \cite{sbert}, introduced a new paradigm that excels at understanding meaning and context. These semantic models can successfully bridge the semantic gap where lexical methods fail. However, these models introduce their own critical failure modes. The first is lexical fragility: their knowledge is based on a fixed vocabulary, making them vulnerable to out-of-vocabulary misspellings. The second is semantic over-reliance, where models, in their focus on meaning, incorrectly group distinct items. A pure semantic search might identify \textit{"Apple iPhone 14 Pro"} and \textit{"Samsung Galaxy S23 Ultra"} as highly similar, ignoring the critical, lexically-distinct keywords \textit{"Apple"} and \textit{"Samsung"} that are essential for a correct match. This ambiguity is the key failure point for fully unsupervised methods, like \cite{zeroer, christen2023, christen2025}, which can easily mistake highly similar but distinct records. This inability to distinguish between high similarity and true identity necessitates sophisticated, often supervised, matching logic.

The advent of these pre-trained models has shifted the state-of-the-art to the supervised technique of fine-tuning in order to overcome semantic over-reliance, leading to two dominant architectural paradigms: bi-encoders and cross-encoders. The bi-encoder architecture \cite{zeakis2023, li2021bert, scblock, autoblock, bert-er, karapiperis2025experimental,lsblock, blink} generates a fixed-size vector embedding for each record independently. This is highly scalable, as embeddings can be pre-computed and stored in an Approximate Nearest Neighbor (ANN) index for efficient, sub-linear similarity search, making it ideal for large-scale blocking. The cross-encoder architecture, popularized by Ditto \cite{ditto} and experimentally studied in \cite{ertransformer}, is "accurate-but-slow". It achieves superior accuracy by concatenating the text of a record pair into a single input, allowing a model to perform deep, token-level interaction. However, this accuracy comes at a prohibitively high computational cost. A cross-encoder cannot pre-compute embeddings for indexing and is thus infeasible for large-scale applications without a preceding blocking step by a bi-encoder.

While these advanced supervised models have pushed the state-of-the-art in matching accuracy, they all depend on a critical resource that is expensive, and often infeasible, to obtain : a large set of high-quality, labeled training data. To solve this labeling bottleneck, Active Learning (AL) has emerged as a promising semi-supervised approach. AL aims to minimize human effort by having a model actively query an \textit{Oracle} (a human expert) to label only the most informative or representative pairs. However, naive Active Learning pipelines introduce new scalability challenges. Running predictions on an entire unlabeled pool of millions of candidate pairs can lead to significant memory and compute bottlenecks. Furthermore, a simplistic query strategy--such as focusing only on informative pairs or representative pairs--can lead to slow convergence or a biased model \cite{mourad2024erabqs}.

Existing AL approaches present a clear trade-off between semantic power and computational cost. On one hand, traditional AL methods \cite{qian2017active, al-crowd1, mourad2024erabqs} are not based on deep learning. Some learn human-readable rules but lack the power to capture deep semantic meaning. Others, while pioneering key AL concepts, are computationally expensive and were designed for traditional classifiers like SVMs, Decision Trees, or Random Forests that use simple, hand-crafted lexical features (e.g., Levenshtein or Jaccard). These methods either suffer from complex, slow query strategies or are not equipped to handle the semantic ambiguity of modern datasets. On the other hand, modern deep AL systems leverage powerful transformer-based models but introduce new, massive computational bottlenecks. DIAL \cite{dial}, for example, jointly learns a blocker and matcher but incurs a high computational cost by re-training both components from scratch in every AL round. AL-Risk \cite{alsampling} avoids this but introduces its own high complexity by requiring a separate risk model and solving an NP-hard selection problem in each iteration.

The related work shows a clear gap: existing AL methods are either (1) semantically weak and built on traditional lexical features or (2) semantically powerful but computationally expensive and complex. Our proposed system, ALER (Active Learning for Entity Resolution), is a novel solution designed to bridge this gap, achieving the semantic power of deep learning without the high computational cost. We accomplish this through four key architectural innovations:

\begin{itemize} 

\item \textbf{Scalable Frozen Architecture:} We propose a resource-efficient pipeline that eliminates training and memory bottlenecks by combining \textit{one-time} SBERT encoding with a "divide-and-conquer" partitioning strategy. Instead of computationally expensive fine-tuning, ALER iteratively trains a lightweight classifier on static, semantically partitioned chunks (via K-Means). This design accelerates iterations by orders of magnitude and allows large-scale processing within standard memory constraints without sacrificing semantic expressiveness.

\item \textbf{Hybrid Query Strategy:} We introduce a balanced query strategy that simultaneously targets "confused" pairs (uncertainty sampling) to refine the decision boundary and "confident" pairs (exploitation) to identify and correct high-confidence errors. This dual approach prevents the model from becoming confidently wrong and ensures robust convergence even in domains with high semantic ambiguity.

\item \textbf{Two-Stage Cascade Classifier:} We propose a cascaded resolution architecture that balances speed and precision. A fast recall-oriented model filters the full candidate space, while a smart precision-oriented model—trained on lexical features like Jaro-Winkler—selectively re-ranks the pairs. This allows ALER to achieve state-of-the-art accuracy on heterogeneous data without incurring the latency of a full cross-encoder scan.
\end{itemize}

The remainder of this paper is organized as follows. Section \ref{sec:related} reviews the related work on deep learning for ER and existing active learning strategies, while in Section \ref{sec:problem}, we formally define the problem. Section \ref{sec:evaluation} presents our experimental methodology, dataset descriptions, and a comprehensive evaluation of ALER against state-of-the-art baselines. Finally, Section \ref{sec:conclusions} concludes the paper and outlines directions for future work.
\

\section{Related Work}
\label{sec:related}
The application of deep learning to ER has evolved from early pioneering works, such as DeepMatcher \cite{deepmatcher} and DeepBlocker \cite{deepblocker}, which explored recurrent neural networks and attention mechanisms, to the now-dominant paradigm of fine-tuning large pre-trained language models. 

Recent frameworks that have explored deep learning for ER tasks are Sudowoodo \cite{sudowoodo} and EMBER \cite{ember}, but with significant trade-offs. Sudowoodo presents a multi-purpose, self-supervised framework that uses contrastive learning to learn similarity-aware data representations from unlabeled data, which can be used for blocking or few-shot fine-tuning. Its main weakness in an unsupervised setting is a complex pseudo labeling step, which requires prior knowledge of the positive label ratio and is sensitive to noise and bias. In contrast, EMBER is a supervised "no-code" system that learns task-specific embeddings for similarity-based keyless joins using a contrastive triplet loss. Its primary weakness is this fundamental reliance on labeled examples, as its performance "dramatically declines" without this supervision; it also struggles when data is extremely sparse or lacks sufficient context.

A review of the AL techniques reveals a clear and persistent trade-off between semantic power and computational feasibility. Many foundational AL systems are computationally expensive and were designed for traditional classifiers and lexical features, while modern deep methods that leverage powerful Transformer models introduce high computational cost bottlenecks. Specifically, the state-of-the-art in AL in the context of ER includes the following works: 

Jain et al. \cite{dial} present a system that jointly learns an integrated cross-encoder matcher and a blocker (bi-encoder committee). However, it incurs high computational cost, because it re-trains both the matcher and the entire blocker committee from scratch in each AL round. 

Nafa et al. \cite{alsampling} propose an AL strategy that uses a separate risk model to find pairs with a high chance of being mispredicted. Its main weakness though is its high complexity. It requires training an additional complex model, and the core selection process is an NP-hard weighted K-medoids problem.

Mourad et al. \cite{mourad2024erabqs} propose an AL strategy to balance exploration (representativeness) and exploitation (informativeness) . Its primary weakness is that it only uses a random forest classifier trained on pure traditional lexical handcrafted features (e.g., Levenshtein, Jaccard). Its representativeness calculation also has a high computational complexity that must be run in each iteration.

Qian et al. \cite{qian2017active} learn human-readable ER rules. Its primary weakness though is that it is a traditional, rule-based system that does not use deep learning, making it less powerful than modern models at capturing semantic meaning.

Mozafari et al. \cite{al-crowd1} propose scaling crowdsourcing by using "black-box" AL algorithms, which are based on nonparametric bootstrap to minimize the number of questions asked to the crowd. The primary weakness is the high computational overhead of these bootstrap-based methods, which require many  model re-trainings in each iteration. Furthermore, their experiments are solely  based on traditional classifiers like linear SVMs and decision trees using simple lexical features.
\vspace{-0.2cm}

\section{Problem Definition}
\label{sec:problem}
Let $\mathcal{R}$ and $\mathcal{S}$ be two large collections of records. Our goal is to find the set that has as many as possible matching pairs $\mathcal{D} = \{(r, s) \mid r \in \mathcal{R}, s \in \mathcal{S}, r \equiv s\}$, where $\equiv$ denotes that $r$ and $s$ refer to the same real-world entity. This resolution task faces two primary scalability challenges:
\begin{enumerate}
    \item \textbf{Computational Complexity:} The $O(|\mathcal{R}| \times |\mathcal{S}|)$ search space is computationally infeasible.
    \item \textbf{Label Scarcity:} Supervised models require large quantities of accurately labeled training data, which are either expensive or impossible to obtain.
\end{enumerate}
We assume an Oracle $\mathcal{O}(r, s) \rightarrow y \in \{0, 1\}$ that provides a perfect label $y$ for any queried pair and a total labeling budget in terms of the number of submitted queries to $\mathcal{O}$.

\paragraph{Evaluation Metrics}
We evaluate our model using the F1-score ($2 \cdot (P \cdot R) / (P + R)$), which is the harmonic mean of precision $(P)$ and recall $(R)$. Precision measures the fraction of predicted matches that are correct, while recall measures the fraction of all true matches that were successfully found by the classifier. Additionally, we report blocking recall to evaluate the quality of the candidate generation phase; this metric specifically measures the percentage of true matches that survive the blocking step and are thus available for classification. Formally, we define our AL objective as the following optimization problem:

\begin{problem}
\label{problem}
Build a training set $\mathcal{G} \subset \mathcal{R} \times \mathcal{S}$ that is as diverse as possible, given the constraint of the labeling budget, such that a classifier trained on $\mathcal{G}$ maximizes the F1-score on the final resolution task between $\mathcal{R}$ and $\mathcal{S}$ executed in $O(|\mathcal{R}| \log(|\mathcal{S}|))$ time.
\end{problem}

The diversity of $\mathcal{G}$ will produce a classifier that generalizes well to the entire $\mathcal{R} \times \mathcal{S}$ space, thus achieving a higher F1-score. To address Problem \ref{problem}, we employ a scalable, memory-efficient AL pipeline described in the next section.

\section{ALER}
\label{sec:aler}
We propose ALER (Active Learning for Entity Resolution), a scalable pipeline designed to solve the dual challenges of computational bottlenecks and label scarcity. The architecture is a partitioned AL loop that trains a two-stage cascade matcher, which learns from iteratively-acquired labeled pairs. Algorithm \ref{alg:aler_training} summarizes the key phases of the training phase of the ALER pipeline.

The bi-encoder $E$ maps any record $r$ (and $s$) to a static embedding $\mathbf{v}_r = E(r) \in \mathbb{R}^d$ (and $\mathbf{v}_s = E(s) \in \mathbb{R}^d$). Next, we build a Hierarchical Navigable Small Worlds (HNSW)\cite{hnsw} index $\mathcal{I_{\mathbf{V}_\mathcal{S}}}$ on the embeddings $\mathbf{V}_\mathcal{S}$ of $\mathcal{S}$ (Lines \ref{alg1:emb}--\ref{alg1:ind}). An HNSW index is specifically designed to resolve queries in logarithmic, $O(\log{|\mathbf{V}_\mathcal{S}}|)$, time.

A key bottleneck in AL is the creation of a candidate unlabeled pool of records, which should be submitted to $\mathcal{O}$. To solve this, we first select a small, representative sample $\mathbf{V}_{\mathcal{R}s} \subset \mathbf{V}_\mathcal{R}$ using simple random sampling. Subsequently, we partition $\mathbf{V}_{\mathcal{R}s}$ using K-Means in order to create $N$ semantically coherent chunks (Line \ref{alg1:part}): $ \{\mathcal{T}_1, \dots, \mathcal{T}_N\} = \text{K-Means}(\{\mathbf{v}_r\}_{r \in \mathbf{V}_{\mathcal{R}s}}, N)$.

We select $N$ to scale logarithmically with the size of the training sample $\mathbf{V}_{\mathcal{R}s}$(Line \ref{alg1:N}). This strategy avoids the pitfalls of over-partitioning, which can fragment coherent groups and force the AL loop to redundantly re-learn similar decision boundaries across multiple chunks. We empirically validate this design choice and demonstrate the impact of varying $N$ on F1-score in Section \ref{sec:evaluation}.

Then, by querying index $\mathcal{I}_{\mathbf{V}_\mathcal{S}}$ for the top-$k$, e.g., $k=10$, candidates of each embedding $\textbf{v}_r$ in each $\mathcal{T}_i$, we induce $N$ candidate pool partitions: $\mathcal{C}_i = \bigcup_{ \textbf{v}_r \in \mathcal{T}_i} \text{q}(\mathbf{v}_r, \mathcal{I}_{\mathbf{V}_\mathcal{S}}, k)$. This partitioning of the candidate pool is the core of our scalable, memory-efficient approach. 

To ensure a fast, consistent, and unbiased evaluation benchmark across all iterations, we create a single, fixed validation set $\mathcal{V}$. This set is globally representative because it is built by sampling a small, e.g., $0.1 \times |\mathcal{C}_i|$, fixed number of candidate pairs from each $\mathcal{C}_i$ during a preliminary pass. This stratified validation set, is then re-used to evaluate the F1-score of the model trained in every subsequent iteration, providing a stable and reliable metric to measure the model's general improvement. Lines \ref{alg1:loop1s}--\ref{alg1:loop1e} outline the loop that creates the candidate pools $\mathcal{C}_i$ and $\mathcal{V}$. ALER also initializes its training set $\mathcal{G}$ with a small budget $B_{\textit{seed}}$ of labels using $\mathcal{O}$ (Line \ref{alg1:seed}).

The pipeline then runs a mini AL loop on each pool $\mathcal{C}_i$. In each iteration, we use a labeling budget $B$ to query $\mathcal{O}$ for $B$ new match statuses. The AL loop runs for at most $I_{\text{max}}$ iterations or terminates early if the validation F1-score fails to improve for a specified number of consecutive iterations, defined by our \textit{patience} parameter (explained in Section \ref{sec:key_alg}), as Line \ref{alg1:early} suggests. In each iteration, ALER performs the following steps:
\vspace{-1mm}
\begin{enumerate}   
    \item Trains a lightweight Siamese MLP (Multi-Layer Perceptron) classifier $M_R$ (Line \ref{alg1:train}), detailed in Section \ref{sec:key_alg}. For each embedding pair $(\mathbf{v}_r, \mathbf{v}_s)$ in $\mathcal{G}$, ALER first creates an interaction vector $I_{\textbf{v}_r, \textbf{v}_s} = (\mathbf{v}_r, \mathbf{v}_s, |\mathbf{v}_r - \mathbf{v}_s|, \mathbf{v}_r \cdot \mathbf{v}_s)$. This vector, which includes the original embeddings, their element-wise absolute difference, and their element-wise product, is then fed into the MLP. This structure allows the model to learn complex, non-linear patterns indicating a match, and is significantly faster to train than re-fine-tuning the full SBERT model in each iteration.
    \item Predicts on that partition's unlabeled candidate pool $\mathcal{C}_i$, generating a probability score $\theta \in [0, 1]$ for each embedding pair indicating the likelihood of a match (Line \ref{alg1:predict}). To generate these predictions, the classifier creates the interaction vectors for the embedding pairs in each $\mathcal{C}_i$.
    \item Selects the $B$ most valuable pairs by combining two complementary approaches: exploitation, which selects the most confident pairs, e.g., $\theta \approx 1.0$) and exploration, which selects the most confused or uncertain pairs, e.g., $\theta \approx 0.5$). This is a critical design choice for efficiency. The most confused pairs, which lie near the $0.5$ decision boundary, are the most informative for teaching the model where to draw the line between a match and a non-match. The most confident pairs are queried to achieve two goals: to quickly find new, rare true positives, and more importantly, to identify and correct the model's most critical errors---the high-confidence false positives. This strategy ensures the model simultaneously refines its decision boundary on the hardest cases and corrects its most significant mispredictions in each iteration. This pair selection step is illustrated in Line \ref{alg1:select}. A key methodological choice is that ALER relies on the pre-trained SBERT model to already know how to handle the easy negatives, allowing the AL loop to focus the entire labeling budget on teaching the MLP how to solve the hard, ambiguous cases.    
    \item Gathers perfect labels from $\mathcal{O}$ for the selected pairs and adds them to the training set $\mathcal{G}$ (Line \ref{alg1:fuse}).
\end{enumerate}

\begin{algorithm}[t]
    \caption{ALER: Partitioned Active Learning (Training Phase)}
    \label{alg:aler_training}
    \footnotesize
    \begin{algorithmic}[1]
        \State \textbf{Input:} Datasets $\mathcal{R}, \mathcal{S}$; Oracle $\mathcal{O}$; SBERT Encoder $E$; Neighbors $k$; Budgets $B_{\text{seed}}, B$; Maximum Iterations $I_{\text{max}}$; Patience $p$; Sample proportions $g_s$ and $g_v$.
        \State \textbf{Output:} Classifiers $M_R, M_P$; Thresholds $\theta_R, \theta_P$; Index $\mathcal{I}_{\mathbf{V}_\mathcal{S}}$.

        \State $\mathbf{V}_\mathcal{R} \leftarrow E(\mathcal{R})$ \Comment{Generate static embeddings once} \label{alg1:emb}        
        \State $\mathbf{V}_\mathcal{S} \leftarrow E(\mathcal{S})$ 
        \State $\mathcal{I}_{\mathbf{V}_\mathcal{S}} \leftarrow \text{BuildIndex}(\mathbf{V}_\mathcal{S})$ \Comment{Build a global index on $\mathbf{V}_\mathcal{S}$} \label{alg1:ind}
        
        \State $\mathbf{V}_{\mathcal{R}s} \leftarrow \text{Sample}(\mathbf{V}_\mathcal{R},  g_s|\mathbf{V}_\mathcal{R}| )$ \Comment{Create a small, representative sample} \label{alg1:sample}
        \State $N \leftarrow \lceil \log_{10}(\mathbf{V}_{\mathcal{R}s}) \rceil $ \Comment{$N$ is scaled logarithmically with the size of $\mathbf{V}_{\mathcal{R}s}$} \label{alg1:N}
        \State $\{\mathcal{T}_1, \dots, \mathcal{T}_N\} \leftarrow \text{K-Means}(\mathbf{V}_{\mathcal{R}s}, N)$ \Comment{Partition the sample} \label{alg1:part}
         
        \State $\mathcal{V} \leftarrow \emptyset$
        \For{$i \leftarrow 1$ \textbf{to} $N$} \label{alg1:loop1s}
            \State $\mathcal{C}_i \leftarrow \bigcup_{\mathbf{v}_r \in \mathcal{T}_i} \text{q}(\mathbf{v}_r, \mathcal{I}_{\mathbf{V}_\mathcal{S}}, k)$ \Comment{Get candidates for chunk $\mathcal{T}_i$} \label{alg1:v1}
            \State $\mathcal{V} \leftarrow \mathcal{V} \cup \mathcal{O} (\text{Sample}(\mathcal{C}_i, g_v|\mathcal{C}_i|))$ \Comment{Sample and label a small quantity of pairs from each $\mathcal{C}_i$ to build the global validation set $\mathcal{V}$} \label{alg1:v3}            
        \EndFor \label{alg1:loop1e}
        
        \State $\mathcal{G} \leftarrow \mathcal{O}(\text{Sample}(\mathcal{C}_1, B_{\text{seed}}))$ \label{alg1:seed} \Comment{Create the initial seed training set by randomly sampling $B_{\text{seed}}$ pairs from $\mathcal{C}_1$.} 
        
        \For{$i \leftarrow 1$ \textbf{to} $N$} \Comment{Iterate over each chunk}

            \For{$j \leftarrow 1$ \textbf{to} $I_{\text{max}}$} \Comment{Run "mini" AL loop}               
                \State $(M_R, f1) \leftarrow \text{TrainRecallClassifier}(\mathcal{G}, \mathcal{V})$ \label{alg1:train}

                \If{\text{EarlyStoppingCriteriaMet}(f1, p)} \textbf{break} \EndIf \Comment{ Terminate if the F1-score on $\mathcal{V}$ fails to improve by a minimum threshold for a pre-defined number of patience $p$ consecutive iterations.} \label{alg1:early}
                
                \State $P \leftarrow M_R.\text{predict}(\mathcal{C}_i)$ \Comment{Predict on chunk's pool}
                \label{alg1:predict}
                \State $Q \leftarrow \text{SelectPairs}(P, B)$ \Comment{Select the $B$ most "confused" and "confident" pairs} \label{alg1:select}               
                \State $\mathcal{G} \leftarrow \mathcal{G} \cup  \mathcal{O}(Q)$ \Comment{Fuse clean labels into $\mathcal{G}$} \label{alg1:fuse}
            \EndFor            
        \EndFor
        
        \State $(M_R, \theta_R, f1) \leftarrow \text{TrainRecallClassifier}(\mathcal{G}, \mathcal{V})$ \label{alg1:trainR}
        \State $(M_P, \theta_P, f1) \leftarrow \text{TrainPrecisionClassifier}(\mathcal{G}, \mathcal{V})$ \label{alg1:trainP}
        
        \State \Return $M_R, M_P, \theta_{P}, \theta_{R}, \mathcal{I}_{\mathbf{V}_\mathcal{S}}$
    \end{algorithmic}
\end{algorithm}

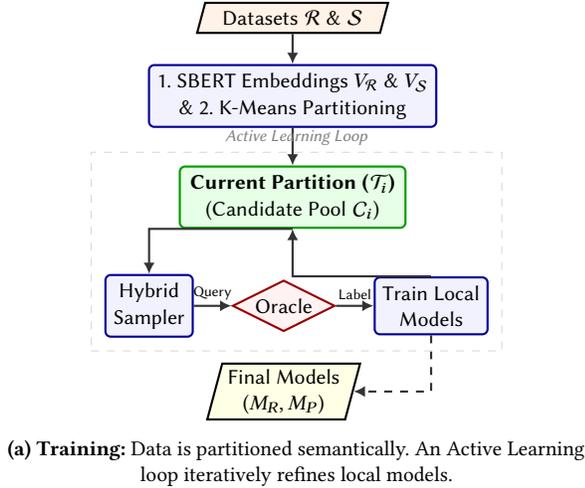
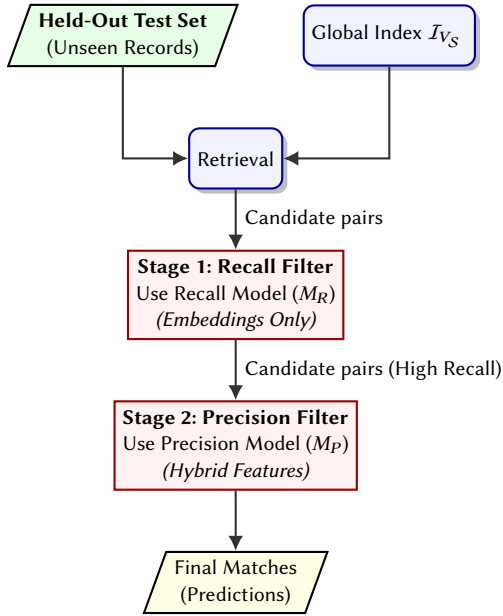
\begin{figure}[h]
    \centering
    \begin{subfigure}[b]{0.45\textwidth}

\centering
\begin{tikzpicture}[
    node distance=0.6cm and 0.6cm,
    font=\sffamily\small,
    >={Latex[width=1.5mm,length=1.5mm]},
    box/.style={rectangle, draw=blue!60!black, fill=blue!5, thick, rounded corners=2pt, minimum height=0.5cm, align=center, inner sep=3pt},
    data/.style={trapezium, trapezium left angle=75, trapezium right angle=105, draw=black, fill=orange!10, thick, align=center, minimum height=0.4cm, inner sep=2pt},
    human/.style={diamond, draw=red!60!black, fill=red!5, thick, aspect=2, align=center, inner sep=1pt},
    arrow/.style={->, thick, draw=gray!40!black}
]

    \node[data] (input) {Datasets $\mathcal{R}$ \& $\mathcal{S}$};
    \node[box, below=0.4cm of input] (prep) {1. SBERT Embeddings $V_{\mathcal{R}}$ \&  $V_{\mathcal{S}}$ \\ \& 2. K-Means Partitioning};
    \draw[arrow] (input) -- (prep);

    \node[box, below=0.5cm of prep, fill=green!10, draw=green!60!black, minimum width=3cm] (partition) {\textbf{Current Partition ($\mathcal{T}_i$)} \\ (Candidate Pool $\mathcal{C}_i$)};
    \draw[arrow] (prep) -- (partition);

    \node[box, below left=0.6cm and -0.2cm of partition] (sampler) {Hybrid \\ Sampler};
    \node[human, right=0.5cm of sampler] (oracle) {Oracle};
    \node[box, right=0.5cm of oracle] (train) {Train Local \\ Models};

    \draw[arrow] (partition.south) -| (sampler.north);
    \draw[arrow] (sampler) -- node[above, scale=0.7] {Query} (oracle);
    \draw[arrow] (oracle) -- node[above, scale=0.7] {Label} (train);
    \draw[arrow] (train.north) -| node[right, scale=0.7] {} (partition.south);

    \node[data, below=0.4cm of oracle, fill=yellow!10] (final) {Final Models \\ ($M_R, M_P$)};
    \draw[arrow, dashed] (train) |- (final);

    \node[draw=gray!30, dashed, inner sep=5pt, fit=(partition) (sampler) (oracle) (train), label={[scale=0.8, gray]above:\textit{Active Learning Loop}}] {};

\end{tikzpicture}
\caption{\textbf{Training:} Data is partitioned semantically. An Active Learning loop iteratively refines local models.}
\label{fig:al_training}
\end{subfigure}
\hfill
  \begin{subfigure}[b]{0.45\textwidth}
\centering
\begin{tikzpicture}[
    node distance=1.0cm and 1.0cm,
    font=\sffamily\small,
    >={Latex[width=2mm,length=2mm]},
    process/.style={rectangle, draw=blue!60!black, fill=blue!5, thick, rounded corners, minimum height=0.8cm, align=center, drop shadow},
    data/.style={trapezium, trapezium left angle=70, trapezium right angle=110, draw=black, fill=orange!10, thick, align=center, minimum height=0.7cm},
    filter/.style={rectangle, draw=red!60!black, fill=red!5, thick, minimum width=2.5cm, minimum height=1cm, align=center},
    arrow/.style={->, thick, draw=gray!40!black}
]

    \node[data, fill=green!10] (testset) {\textbf{Held-Out Test Set} \\ (Unseen Records)};
    \node[process, right=1cm of testset] (index) {Global Index $\mathcal{I}_{V_{\mathcal{S}}}$ };

    \node[process, below=0.8cm of testset, xshift=1.5cm] (retrieval) {Retrieval};
    \draw[arrow] (testset) |- (retrieval);
    \draw[arrow] (index) |- (retrieval);

    \node[filter, below=0.8cm of retrieval] (stage1) {\textbf{Stage 1: Recall Filter} \\ Use Recall Model ($M_R$) \\ \textit{(Embeddings Only)}};
    \draw[arrow] (retrieval) -- node[right] {Candidate pairs} (stage1);

    \node[filter, below=0.8cm of stage1] (stage2) {\textbf{Stage 2: Precision Filter} \\ Use Precision Model ($M_P$) \\ \textit{(Hybrid Features)}};
    \draw[arrow] (stage1) -- node[right] {Candidate pairs (High Recall)} (stage2);

    \node[data, below=0.8cm of stage2, fill=yellow!10] (results) {Final Matches \\ (Predictions)};
    \draw[arrow] (stage2) -- (results);


\end{tikzpicture}
\caption{\textbf{Two-Stage Resolution on Held-Out Test Set:} To ensure rigorous evaluation, the learned models are applied to a strictly held-out test set. Stage 1 acts as a learnable blocker, while Stage 2 applies fine-grained hybrid features only to difficult candidates.}
\label{fig:resolution}
\end{subfigure}
\caption{The ALER pipeline}
\label{fig:pipeline}
\end{figure}

A final re-training step (Line \ref{alg1:trainR}) is executed to incorporate the last batch of labels. This prevents model staleness and ensures $M_R$ utilizes the complete set $\mathcal{G}$ for the resolution phase. Figure \ref{fig:al_training} illustrates the workflow of the training phase.

The above steps fuse the most valuable labels from all semantic chunks into one diverse high-quality training set $\mathcal{G}$, which is used to train our precision-oriented Siamese MLP classifier $M_P$ (Line \ref{alg1:trainP}). This training process uses the computationally slower lexical features for each embedding pair $(\mathbf{v}_r, \mathbf{v}_s)$ in order to formulate each lexical feature vector $H_{r,s} =(\mathbf{v}_r, \mathbf{v}_s,  |\mathbf{v}_r - \mathbf{v}_s|, \mathbf{v}_r \cdot \mathbf{v}_s), f(r, s))$. In these vectors, component $f$ includes structural features, such as Jaro-Winkler similarities, computed on key attributes (e.g., \texttt{title}, \texttt{name}), which possess high discriminative power but are susceptible to subtle lexical variations such as typos or abbreviations. While these features provide a complementary syntactic signal to the model's semantic understanding, computing them is a more expensive operation that requires fetching the original text for specific string attributes and executing CPU-bound string similarity calculations.

Upon completion of training, each classifier ($M_R$ and $M_P$) generates optimal probability thresholds $\theta_R$ and $\theta_P$\footnote{These thresholds are not similarity (or distance) thresholds of the corresponding embedding pairs.}. After each model is trained, it generates probability scores on the held-out validation set. These probabilities, along with the true labels are used to generate a full precision-recall curve. ALER then computes the F1-score for all possible thresholds along this curve and selects the threshold that corresponds to the maximum F1-score. The chosen thresholds are then applied as decision boundaries in the final resolution cascade.  

The final resolution task is a two-stage cascade designed to maximize both scalability and accuracy. This cascade uses the fast, recall-oriented model $M_R$ as a broad filter to reduce the search space, and then deploys the precision-oriented model $M_P$ only on the high-potential candidates that pass the first stage. 

For the first stage, to ensure a strict held-out evaluation and avoid data leakage, we enforce a record-level exclusion strategy. Specifically, for every candidate pair $(\textbf{v}_r, \textbf{v}_s)$ retrieved by querying $\mathcal{I}_{\mathbf{V}_\mathcal{S}}$, we compute the interaction feature vector $I_{\textbf{v}_r, \textbf{v}_s}$ only if the query record $\textbf{v}_r$ does not appear in any instance within the training set $\mathcal{G}$ or the validation set $\mathcal{V}$. These vectors are then fed in batches into $M_R$ to predict their match status. The output is a probability score for the match status of every such candidate pair. We then filter this list, keeping only the embedding pairs whose probability exceeds the model's optimized recall threshold $\theta_R$.

The second stage operates only on the smaller, high-recall set of pairs that passed the first stage. For each of these hard pairs, we compute its lexical feature vector $H_{r,s}$. These new, richer vectors—combining embedding interactions with lexical features—are fed in batches into $M_P$, enabling it to accurately distinguish difficult false positives. The resulting probabilities are filtered using the precision model's own optimal threshold $\theta_P$ to produce the final, high-precision, high-recall list of matches. Figure \ref{fig:resolution} provides a breakdown of the resolution task.

\subsection{Key Design Choices}
\label{sec:key_alg}

A purely greedy early-stopping mechanism, which halts training after a single iteration of non-improvement of F1-score, is highly susceptible to sampling variance from the validation set. A single noisy F1-score could prematurely terminate the entire AL loop. To make our pipeline more robust, we implement a patience counter. The loop is only terminated if the F1-score on the fixed validation set fails to improve by a minimum threshold, e.g., $0.05$, for \textit{patience} consecutive iterations. This mechanism ensures that the model has a chance to recover from minor, random fluctuations in the validation metric, making the training process more stable and reliable.

We apply K-Means to sample $\mathbf{V}_{\mathcal{R}s}$ to partition it into $N$ semantically coherent chunks to solve both the memory bottleneck and improve label efficiency. Instead of running one general-purpose AL loop, our pipeline runs $N$ mini loops, one for each specialized chunk (e.g., "TVs", "laptops"). This forces the model to focus its limited labeling budget on the most valuable and difficult pairs within each distinct semantic broad area, ensuring our final fused training set $\mathcal{G}$ is both highly diverse and rich with informative hard-negatives. Crucially, this stratified approach mitigates the bias inherent in random sampling; by dedicating a specific loop to smaller chunks, we ensure that the model learns to resolve entities even in underrepresented domains, which would otherwise be ignored by a monolithic global loop.

The labeling budget represents the total, fixed number of pairs we are willing to "pay" our human expert $\mathcal{O}$ to manually label. In our pipeline, this budget is split into two parts: a very small, one-time cost $B_{\textit{seed}}$ (e.g., 100 pairs) needed to train the very first seed model, and the main budget $B$, which is spent iteratively (e.g., $200 \times 5 \text{ iterations}$). The central goal of our AL pipeline is to be "label-efficient" -- that is, to use this limited budget as intelligently as possible by querying $\mathcal{O}$ for only the most valuable pairs. This allows the system to achieve a much higher F1-score than if it were trained on the same number of randomly sampled pairs.

Both $M_R$ and $M_P$ share the same underlying Siamese Multi-Layer Perceptron (MLP) architecture. The network is designed as a lightweight, feed-forward neural network composed of two hidden layers. The first dense layer consists of $128$ neurons with ReLU activation, followed by a dropout layer  to prevent overfitting. This is connected to a second dense layer of 64 neurons and a subsequent dropout layer. The final output layer contains a single neuron with a sigmoid activation function, producing a probability score $\theta \in [0, 1]$ indicating the likelihood of a match. Training is performed with a batch size of $64$ for a maximum of $15$ epochs.

Resolving the match status of the most confident pairs and including them in $\mathcal{G}$ refines the decision boundary for difficult and ambiguous cases. For instance, consider a pair that exhibits high similarity but is not a match, e.g., a pair of different device models from the same manufacturer. Querying the most confident pairs is particularly effective here, as the model often mistakenly assigns a high probability score to these false positives due to their overwhelming semantic overlap. By actively exposing and correcting these predictions, we force the classifier to reject such instances, thereby significantly improving precision without sacrificing recall.

\begin{figure*}[t] 	
 		\includegraphics[width=1\linewidth]{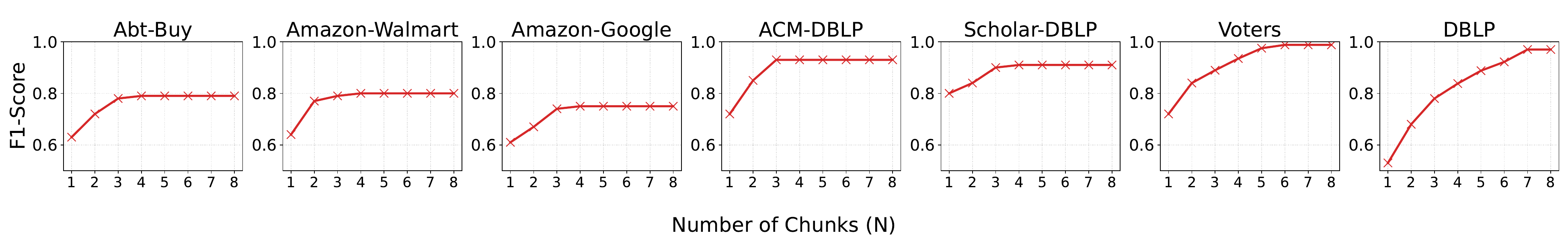}
		\caption{Effect of Chunk Count $(N)$ on F1-Score $(B=300)$}    
		\label{fig:cluster}	            
\end{figure*}

\section{Experimental Evaluation}
\label{sec:evaluation}
We designed a series of experiments across nine diverse datasets, encompassing real-world benchmarks and large-scale semi-synthetic datasets. We used product catalogs (Abt-Buy\footnote{\label{note1}\url{https://old.dbs.uni-leipzig.de/research/projects/object_matching/benchmark_datasets_for_entity_resolution}}, Amazon-Walmart\footnote{\url{https://hpi.de/naumann/projects/repeatability/datasets/amazon-walmart-dataset.html}}, and Amazon-Google\textsuperscript{\ref{note1}}), bibliographic datasets (ACM-DBLP\textsuperscript{\ref{note1}} and Scholar-DBLP\textsuperscript{\ref{note1}}), and datasets spanning the movie and restaurant domains (IMDB-DBPEDIA\footnote{\url{https://zenodo.org/record/8433873/files/data_ea.tar.gz}} and Restaurants\footnote{\url{https://hpi.de/naumann/projects/repeatability/datasets/restaurants-dataset.html}}). Additionally, we included two large-scale datasets: Voters\footnote{\url{https://www.ncsbe.gov/results-data/voter-registration-data}} ($1\text{M}\times1\text{M}$ records) and DBLP\footnote{\url{https://dblp.org/xml}} ($3\text{M}\times3\text{M}$ records). These large-scale datasets are semi-synthetic, created by perturbing the records of a real dataset to generate its counterpart. The selected datasets are all well-established benchmarks for ER in the literature~\cite{webofdatabook, tkde2024, pap16, gaglia1, sudowoodo, autoblock, deepmatcher, deepblocker, ditto, gnem, kar-pvldb, kar2014, lsblock, zeakis2023, deeper, dial, ertransformer, karapiperis2025experimental, flexer, mourad2024erabqs}.

All experiments were conducted on a machine equipped with $80$ GB of system RAM and an NVIDIA GPU with $23$ GB of VRAM. Our implementation uses the HNSW ANN index, offered by FAISS\footnote{\url{https://github.com/facebookresearch/faiss}}, which uses highly-optimized operations for both index construction and sub-linear search time. SBERT encoding, using the 384-dim \texttt{MiniLM-L6-v2} embedding model~\cite{mini}, is a one-time, GPU accelerated batch operation that processes the entire corpus. To ensure the reliability of the experiments, the presented results are the average values from $10$ experimental runs. The size of the sample $\mathbf{V}_{\mathcal{R}s}$ was set to $|\mathbf{V}_{\mathcal{R}s}|=0.2 \times |\mathbf{V}_\mathcal{R}|$.

We evaluate our proposed framework against three state-of-the-art AL baselines for ER, covering diverse query strategies and architectural paradigms. We specifically prioritized AL methods over fully supervised deep learning models, as the latter require massive labeled datasets to converge and are thus inapplicable to the label-scarce regimes targeted by our work. DIAL \cite{dial} is a deep AL system that jointly optimizes a bi-encoder for blocking and a cross-encoder for matching within a single feedback loop, representing the current state-of-the-art in deep ER accuracy. AL-Risk \cite{alsampling} introduces a risk-based query strategy, which trains a separate risk estimation model to select instances that minimize the expected misprediction error, formulating the selection as a weighted K-medoids problem. Finally, ERABQS \cite{mourad2024erabqs} employs a balancing strategy that explicitly weighs "informativeness" (uncertainty) against "representativeness" (density) to guide the learning process, utilizing a Random Forest classifier trained on traditional lexical features. We configured DIAL to jointly learn a blocker and a matcher using a committee of pre-trained RoBERTa~\cite{roberta} models that are re-trained in each AL round, utilizing a budget of $200$ pairs per round for $12$ rounds. We also adapted the AL-Risk query budgets to the dataset sizes (e.g., $200$ to $575$ pairs). Finally, ERABQS was configured with a Random Forest classifier ($100$ trees) using  Levenshtein distances and utilizing an $\epsilon$-decreasing strategy, starting from $\epsilon=1.0$ to balance exploration and exploitation.

\begin{figure*}[t] 	
 		\includegraphics[width=1\linewidth]{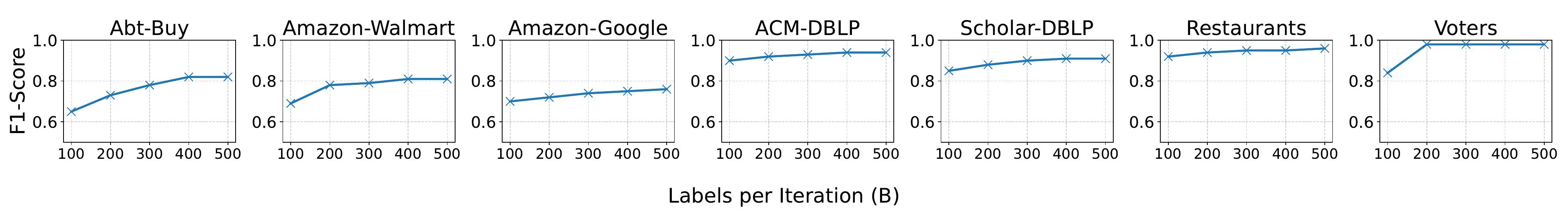}
		\caption{Effect of Budget Size $(B)$ on Final F1-Score (max $8$ Iterations)}   
		\label{fig:ablation}	            
\end{figure*}
\begin{figure*}[t] 	
 		\includegraphics[width=1\linewidth]{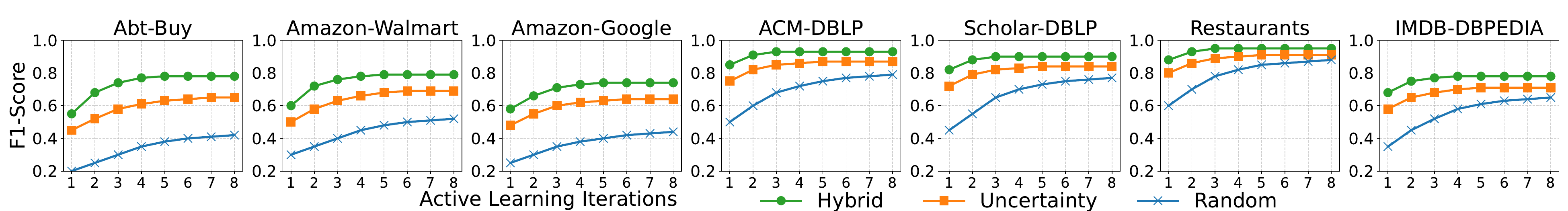}
		\caption{Query Strategy Effectiveness $(B=300)$}    
		\label{fig:query}	 
\end{figure*}

\subsection{Experimental results}
\label{sec:results}
\textbf{Impact of Partition Count ($N$):} In the first experiment, we investigated the effect of varying the number of semantic chunks ($N$) on the final F1-score illustrated in Figure \ref{fig:cluster}. The results reveal a strong, non-linear relationship between the granularity of partitioning and model performance. There is a considerable improvement in F1-scores across all datasets as we increase the number of chunks depending on the size of the sample $\mathbf{V}_{\mathcal{R}s}$, which in turn depends on the size of $\mathbf{V}_\mathcal{R}$. For example, setting $N=3$ Scholar-DBLP sees a gain of $+0.11$ ($0.79 \rightarrow 0.90$), and ACM-DBLP improves by $+0.21$ ($0.72 \rightarrow 0.93$). This suggests that without partitioning, a simple random sample fails to capture the diverse structural patterns (e.g., different citation formats) needed to train a robust model. The gains are also significant for the challenging product datasets. Abt-Buy improves by $+0.15$ ($0.63 \rightarrow 0.78$) and Amazon-Walmart by $+0.15$ ($0.64 \rightarrow 0.79$), when setting $N=3$. On the other hand, the large datasets, Voters and DBLP, reach their performance plateau at $N=5$ and $N=7$, respectively. These values suggest a nearly logarithmic relationship between $N$ and the sample size $|\mathbf{V}_{\mathcal{R}s}|$. Increasing the number of chunks beyond these thresholds yields negligible improvement, a trend consistent with our findings on smaller datasets (e.g., Scholar-DBLP stays flat at $0.91$ and Amazon-Google at $0.74$ for $N \ge 3$). This indicates that a moderate level of partitioning is sufficient to disentangle the semantic space without inducing unnecessary overhead. For the rest of the experiments, we used $N=3$ for the small datasets, and $N=5$ and $N=7$ for Voters and DBLP, respectively.

\textbf{Impact of Labeling Budget ($B$):} Subsequently, we varied the budget size ($B$) from $100$ to $500$. We observed that the number of iterations required to trigger early stopping was inversely proportional to the budget size; while smaller batches (e.g., $B=100$) required an average of $4$ iterations to saturate the model's performance, larger batches (e.g., $B=300$) terminated in approximately $2$ iterations. The results, visualized in Figure \ref{fig:ablation}, demonstrate a robust positive correlation between $B$ and the final F1-score of the two-stage cascade matcher across all seven datasets. The primary trend indicates that increasing $B$ consistently improves model performance, but this improvement follows a steep curve of diminishing returns. The most substantial gains are concentrated in the transition from a highly constrained budget ($B=100$) to a moderate one ($B=200$ or $B=300$). On the Abt-Buy dataset, for instance, increasing the batch size from $100$ to $300$ yields a $0.13$ improvement in F1-score ($0.65 \rightarrow 0.78$), whereas further increasing it from $300$ to $500$ yields a smaller marginal gain of $0.04$. This plateau effect is even more pronounced on Amazon-Walmart, ACM-DBLP, and Scholar-DBLP, where the F1-score remains virtually flat between batch sizes of $400$ and $500$. Notably, the Voters dataset reaches saturation even earlier, showing no performance gain beyond $B=200$. These findings highlight a critical cost-benefit trade-off: practitioners can achieve the vast majority of optimal performance with a moderate budget (e.g., $B=300$), avoiding the $66\%$ cost increase required to reach $B=500$ for often negligible returns.

The results in Figure \ref{fig:ablation} also highlight the pipeline's varying effectiveness across different data types. ALER achieves excellent F1-scores (approaching or exceeding $0.90$) on structured or \textit{cleaner} textual data, as seen in the ACM-DBLP, Scholar-DBLP, Restaurants, and Voters datasets. Furthermore, we observe that the Abt-Buy and Amazon-Google datasets exhibit notably slower convergence than the other benchmarks. While datasets like Voters reach their performance plateau with a labeling budget of only $B=200$, Abt-Buy and Amazon-Google continue to show meaningful performance gains up to $B=400$ and $B=500$, respectively. This slower convergence rate suggests that the model requires a larger density of examples to resolve the higher degree of semantic and structural ambiguity present in these product domains.

\textbf{Effectiveness of Hybrid Query Strategy:} We conducted an ablation study comparing our hybrid query strategy against two baselines: standard uncertainty sampling (selecting only pairs near the decision boundary) and random sampling. In each iteration of the active learning loop, the pairs selected by these strategies are queried against the Oracle to obtain their ground-truth labels. These newly labeled instances are then added to the cumulative training set, which is used to re-train $M_R$ from scratch, thereby progressively refining its decision boundary and correcting its high-confidence errors for the subsequent round. The results, illustrated in Figure \ref{fig:query}, demonstrate a substantial and consistent performance gap across all seven datasets, confirming the superiority of our approach. ALER consistently achieves the highest F1-score and the fastest convergence. On the challenging product datasets like Abt-Buy and Amazon-Walmart, it establishes a dominant lead early in the learning process. For example, on Abt-Buy, our hybrid strategy reaches an F1-score of $0.78$, whereas the uncertainty strategy plateaus at $0.65$—a significant $13$-point gap. This confirms our hypothesis that querying "confident" pairs ($\theta \approx 1.0$) is critical for correcting the model's high-confidence false positives, preventing it from becoming confidently wrong. The uncertainty strategy performs significantly better than random sampling but consistently lags behind the hybrid approach by a wide margin (typically $10-15\%$). While it effectively refines the decision boundary for ambiguous cases, it fails to detect false positives that lie far from the boundary. On Amazon-Google, for instance, uncertainty sampling saturates at an F1-score of $0.64$. As expected, random sampling yields the poorest performance, often trailing the hybrid strategy by $20-30\%$ or more. For instance, on Abt-Buy, it achieves a final F1-score of only $0.42$. This shallow learning curve demonstrates that the vast majority of candidate pairs in the pool are "easy" negatives that provide little informational value, highlighting the absolute necessity of active selection for resolving large-scale entity resolution tasks efficiently.
\begin{table}[t]
\centering
\footnotesize
\caption{End-to-End Computational Cost Breakdown (Emb: Embedding, Idx: Indexing, Rs: Resolution).}
\label{tab:system_cost_compact}
\vspace{-3mm}
\begin{tabular}{lrrrrrrr}
\toprule
\multirow{2}{*}{\textbf{Dataset}} & \multicolumn{2}{c}{\textbf{Setup (s)}} & \multicolumn{2}{c}{\textbf{Mem (GB)}} & \multicolumn{2}{c}{\textbf{Runtime (s)}} & \textbf{Total} \\ \cline{2-7}
 & \textbf{Emb} & \textbf{Idx} & \textbf{Emb} & \textbf{Idx} & \textbf{Loop} & \textbf{Rs} & \textbf{(m)} \\ \hline
\textbf{Voters} & 588 & 189 & 2.76 & 3.89 & 706 & 310 & \textbf{30} \\
\textbf{DBLP} & 1,488 & 580 & 8.66 & 5.24 & 1,286 & 737 & \textbf{62} \\ 
\bottomrule
\end{tabular}
\vspace{-4mm}
\end{table}

\textbf{End-to-End Cost Breakdown:} To demonstrate the practical deployability of our proposed architecture, we report a granular breakdown of the computational resources in Table \ref{tab:system_cost_compact} required for our largest datasets, Voters and DBLP. For Voters, the one-time cost to embed the corpus ($588$ seconds) and build the HNSW index ($189$ seconds) was approximately $13$ minutes. Following this setup, the full AL loop—encompassing inference, query selection, and validation—completed in just $706$ seconds, supporting nearly \num{4000} queries with negligible retrieval latency. The total end-to-end runtime—including bootstrapping, active learning, and final resolution of the unlabeled remainder—was under $30$ minutes. Similarly, for DBLP dataset, the system completed the full end-to-end pipeline in roughly $62$ minutes with a peak memory footprint of $\approx 14$ GB (combining embeddings and index structures). This confirms that the system operates within the resource constraints of a standard workstation even for datasets containing millions of records.

\begin{table}[h]
\centering
\footnotesize
\caption{Detailed Label Budget Breakdown $(B=300)$.}
\label{tab:budget_breakdown}
\vspace{-3mm}
\begin{tabular}{lrrrrr}
\toprule
\textbf{Dataset} & $N$ & $B_{\text{seed}}$ & $|\mathcal{V}|$ & Loop & Total \\
\midrule
Abt-Buy & 3 & 100 & 81 & 1,057 & 1,438 \\
Amazon-Walmart & 3 & 100 & 1,000 & 1,334 & 2,634 \\
Amazon-Google & 3 & 100 & 269 & 1,800 & 2,369 \\
ACM-DBLP & 3 & 100 & 158 & 1,200 & 1,658 \\
Scholar-DBLP & 3 & 100 & 183 & 1,800 & 2,283 \\
Restaurants & 3 & 100 & 29 & 311 & 640 \\
IMDB-DBPEDIA & 3 & 100 & 1,000 & 1,800 & 3,100 \\
Voters & 5 & 100 & 2,000 & 3,900 & 6,400 \\
DBLP & 7 & 100 & 2,000 & 5,100 & 7,800 \\
\bottomrule
\end{tabular}%
\vspace{-3mm}
\end{table}

\begin{table}[ht]
\centering
\footnotesize
\caption{Comparative evaluation of ALER, AL-Risk, DIAL, and ERABQS. (\textbf{R}: Recall, \textbf{P}: Precision, \textbf{F1}: F1-Score, \textbf{Tr}: Training Time, \textbf{Rs}: Resolution Time).}
\label{tab:comparison}
\vspace{-3mm}
\begin{tabular}{llcccrr}
\toprule
\textbf{Dataset} & \textbf{Method} & \textbf{R(\%)} & \textbf{P(\%)} & \textbf{F1(\%)} & \textbf{Tr(s)} & \textbf{Rs(s)} \\
\midrule
\multirow{4}{*}{\shortstack{\textbf{Abt-Buy} }}
 & DIAL & 85.1 & 59.5 & 70.0 & 232 & 42 \\
 & AL-Risk  & 82.5 & 64.0 & 72.1 & 245 & 41 \\
 & ERABQS  & 75.2 & 57.2 & 64.9 & 115 & 18 \\
 & ALER & 91.5 & 68.2 & 78.1 & 47 & 2 \\
\midrule
\multirow{4}{*}{\shortstack{\textbf{Amazon-Walmart} }}
 & DIAL  & 82.0 & 58.1 & 68.1 & 294 & 88 \\
 & AL-Risk  & 80.0 & 62.5 & 70.2 & 311 & 87 \\
 & ERABQS  & 78.0 & 50.1 & 61.2 & 147 & 25 \\
 & ALER  & 93.2 & 66.2 & 79.0 & 63 & 2 \\
\midrule
\multirow{4}{*}{\shortstack{\textbf{Amazon-Google} }}
 & DIAL  & 89.0 & 64.8 & 75.1 & 212 & 32 \\
 & AL-Risk  & 86.5 & 63.0 & 72.9 & 226 & 29 \\
 & ERABQS  & 79.6 & 48.5 & 60.3 & 107 & 16 \\
 & ALER & 92.1 & 59.5 & 74.2 & 44 & 2 \\
\midrule
\multirow{4}{*}{\shortstack{\textbf{ACM-DBLP} }}
 & DIAL  & 91.0 & 92.5 & 91.7 & 299 & 28 \\
 & AL-Risk  & 91.5 & 91.8 & 91.6 & 305 & 24 \\
 & ERABQS  & 89.5 & 91.0 & 90.2 & 160 & 14 \\
 & ALER & 92.5 & 93.5 & 93.2 & 51 & 2 \\
\midrule
\multirow{4}{*}{\shortstack{\textbf{Scholar-DBLP} }}
 & DIAL  & 87.5 & 89.2 & 88.3 & 496 & 256 \\
 & AL-Risk  & 88.0 & 89.5 & 88.7 & 510 & 251 \\
 & ERABQS  & 85.0 & 87.0 & 86.0 & 211 & 45 \\
 & ALER& 89.5 & 90.5 & 90.1 & 83 & 3 \\
\midrule
\multirow{4}{*}{\textbf{Restaurants}} 
 & DIAL  & 92.5 & 94.0 & 93.2 & 195 & 21 \\
 & AL-Risk  & 93.0 & 94.5 & 93.7 & 202 & 19 \\
 & ERABQS & 91.0 & 93.0 & 92.0 & 95 & 12 \\
 & ALER & 94.5 & 95.5 & 95.3 & 41 & 2 \\
\midrule
\multirow{4}{*}{\shortstack{\textbf{IMDB-DBPEDIA} }}
 & DIAL  & 75.5 & 77.5 & 76.5 & 365 & 120 \\
 & AL-Risk  & 78.0 & 80.0 & 79.0 & 382 & 115 \\
 & ERABQS  & 73.0 & 75.0 & 74.0 & 181 & 61 \\
 & ALER & 95.2 & 80.3 & 87.1 & 74 & 3 \\
\midrule
\multirow{4}{*}{\textbf{Voters}} 
 & DIAL  & 85.4 & 92.2 & 88.7 & 3,402 & 1,245 \\
 & AL-Risk & 89.5 & 91.5 & 90.5 & 2,156 & 2,100 \\
 & ERABQS  & 88.6 & 90.1 & 89.3 & 2,223 & 1,818 \\
 & ALER& 98.5 & 99.0 & 98.7 & 1,483 & 310 \\
\midrule
\multirow{4}{*}{\textbf{DBLP}} 
& DIAL & 85.1 & 90.1 & 87.4 & 6,600 & 3,152 \\
& AL-Risk & 86.5 & 89.5 & 88.0 & 4,245 & 3,223 \\
& ERABQS  & 80.2 & 85.0 & 82.4 & 4,523 & 2,824 \\
& ALER & 96.5 & 97.5 & 97.0 & 3,354 & 737\\
\bottomrule
\end{tabular}%
\vspace{-5mm}
\end{table}

\textbf{Comparing with Fine-Tuned Embeddings:} We also compared our frozen architecture against an iterative fine-tuning baseline, where the SBERT encoder was updated using Contrastive Loss for one epoch at each AL iteration. This approach required re-embedding the candidate pool and re-building the HNSW index at every iteration to allow our query strategy to select valid samples based on the updated vector space. Evaluating blocking recall to isolate embedding quality, we found that fine-tuning yielded negligible gains ($<1\%$) while imposing prohibitive computational costs. The optimization step alone, without the re-embedding and re-indexing tasks, introduced a $\sim 500\times$ latency increase ($\approx 5$ seconds for fine-tuning vs. $0.01$ seconds for training the MLP classifier), which remained constant across all datasets due to the fixed labeling budget. However, re-embedding and re-indexing introduced a severe scalability bottleneck on large datasets like Voters and DBLP, adding $13$ and $35$ minutes (Table \ref{tab:system_cost_compact}) of overhead per iteration, respectively. This confirms that frozen embeddings provide sufficient retrieval quality without the infeasible overhead of fine-tuning.

\textbf{Comparing with the Competitors:} Guided by these empirical findings, we adopted the fixed configuration detailed in Table~\ref{tab:budget_breakdown}, which also quotes the exact label consumption for each dataset, for all subsequent experiments to ensure consistent evaluation. Since ALER utilizes semantic partitioning, the size of each local hypothesis space naturally varies based on the data distribution. In partitions with lower candidate density, the number of informative pairs remaining in the unlabeled pool may fall below the maximum batch size $B$. In such scenarios, the system dynamically adapts by querying only the remaining valid pairs, preventing $\mathcal{O}$ from wasting effort on redundant or empty data. Also, in order to prevent budget explosion, we enforced a strict cap on the validation set $\mathcal{V}$: maximum \num{1000} pairs for small/medium datasets and \num{2000} pairs for large-scale benchmarks. We configured the baselines to match these budget ceilings to within $\pm 5\%$ to ensure parity.

Table \ref{tab:comparison} presents a comprehensive comparison of ALER against their competitors. The results demonstrate that ALER consistently achieves superior or highly competitive F1-scores while offering substantial improvements in computational efficiency. On Abt-Buy, ALER achieves an F1-score of $78.1\%$, significantly outperforming DIAL ($70.0\%$), ERABQS ($64.9\%$), and AL-Risk ($72.1\%$). Similarly, on Amazon-Walmart, ALER reaches $79.0\%$, surpassing DIAL by over $10$ percentage points ($68.1\%$) and AL-Risk by nearly $9$ points ($70.2\%$). On the Amazon-Google dataset, ALER remains highly competitive ($74.2\%$) against DIAL ($75.1\%$) and AL-Risk ($72.9\%$), while maintaining a substantial lead over ERABQS ($60.3\%$). On the structured and bibliographic datasets, ALER consistently pushes the state-of-the-art. For ACM-DBLP and Scholar-DBLP, ALER achieves F1-scores of $93.2\%$ and $90.1\%$ respectively, outperforming all baselines, including the strong deep learning baseline AL-Risk ($91.6\%$ and $88.7\%$). The most critical advantage of ALER lies in its computational efficiency. The frozen bi-encoder architecture significantly reduces training time compared to the computationally intensive routines of the baselines. For example, on the Scholar-DBLP dataset, ALER completes its training loop in just $83$ seconds. In contrast, DIAL and AL-Risk require roughly $500$ seconds ($496$ and $510$) and ERABQS requires $211$ seconds—corresponding to speedups of $6\times$ and $2.5\times$, respectively. This advantage is even more pronounced in resolution times, where ALER consistently resolves datasets in single-digit seconds (e.g., $2$ seconds for Abt-Buy). In stark contrast, DIAL's cross-encoder and AL-Risk's complex sampling architecture require significantly more time, taking $42$ seconds and $41$ seconds, respectively, for Abt-Buy. This represents a $20\times$ speedup in resolution latency.

We extended our evaluation to the large-scale regime, testing the systems on the Voters and DBLP datasets. These benchmarks serve as a stress test for system scalability, where traditional AL bottlenecks become prohibitive. ALER demonstrates exceptional robustness maintaining near-perfect accuracy even as the data volume scales into the millions. On the Voters dataset, ALER achieves an F1-score of $98.7\%$, significantly outperforming DIAL ($88.7\%$), ERABQS ($89.3\%$), and the strong deep learning baseline AL-Risk ($90.5\%$). Similarly, on the DBLP dataset, ALER sustains a high F1-score of $97.0\%$, whereas the baselines struggle to maintain performance, with DIAL and AL-Risk projected to drop to $87.4\%$ and $88.0\%$, respectively, and ERABQS falling to $82.4\%$. This confirms that ALER's partitioned training strategy effectively captures the diversity of large-scale data without being overwhelmed by its size. 

The most profound difference at this scale is evident in the drastic reduction of runtime costs. ALER completes its training loop on Voters in $25$ minutes (\num{1483} seconds), including embedding and indexing. In contrast, the baselines are significantly slower: DIAL requires roughly \num{3402} seconds for the equivalent workflow, largely due to the overhead of full model re-training. AL-Risk and ERABQS require \num{2156} and \num{2223} seconds, respectively, hampered by the high complexity of their risk modeling and feature generation steps. The difference in inference time is equally stark. ALER resolves the full DBLP dataset in approximately $12$ minutes ($737$ seconds). DIAL's cross-encoder architecture requires nearly $53$ minutes (\num{3152} seconds) to perform the same task, with AL-Risk (\num{3223} seconds) and ERABQS (\num{2824} seconds) exhibiting similarly high latencies. These results empirically confirm that ALER remains practical and efficient at scales where the compared methods incur prohibitive latency.

\section{Conclusions and Future Work}
\label{sec:conclusions}
In this paper, we presented ALER, a scalable and label-efficient pipeline for ER. We identified that the primary barrier to adopting deep learning for ER is not just the lack of labels, but the computational overhead of existing AL systems. Our work demonstrates that a bi-encoder architecture coupled with a partitioned AL loop, which trains a lightweight MLP classifier, offers a superior trade-off between speed and accuracy. Our evaluation confirms that ALER achieves state-of-the-art F1-scores with as few as 300 labels per iteration, while scaling to million-record datasets where competitors fail.  Future work will extend this partitioned architecture to streaming environments, utilizing incremental clustering to handle continuous data feeds without re-training.
\balance
\bibliographystyle{ACM-Reference-Format}
\bibliography{references}
\end{document}